\documentclass[
 aps, pra,
 amsmath,amssymb,
 10pt,
 final,
tightenlines,
 twoside,
 twocolumn,
 nobibnotes,
nofootinbib,
 superscriptaddress,
showkeys,
 ]
{revtex4-2}

\usepackage[utf8x]{inputenc}
\usepackage[english]{babel}
\usepackage{graphicx}% Include figure files
\usepackage{dcolumn}% Align table columns on decimal point
\usepackage{bm}% bold math
\usepackage{natbib}
\usepackage{float}
\usepackage{hyperref}
\usepackage{slashed}
\usepackage{multirow}
\usepackage{lineno}
\usepackage{booktabs}
\usepackage{silence}
\usepackage{xcolor}
\usepackage{mathtools}

\usepackage{pifont}
\newcommand{\xmark}{\ding{55}}
\begin{document}

\selectlanguage{english}

%\keywords{stars: double or binary---stars: individual: ADS\,48}

%\ydk{}
%\titlerunning{}
%\authorrunning{}
%\toctitle{}
%\tocauthor{}

\title{Exotic and physics-informed support vector machines for high energy physics}

\author{\firstname{A.}~\surname{Ramirez-Morales}}
 \email{andres.ramirez@fisica.uaz.edu.mx}
\affiliation{Facultad de Física, Universidad Autónoma de Zacatecas, Apartado Postal C-580, 98060 Zacatecas, M\'exico}

  \author{\firstname{A.}~\surname{Gutiérrez-Rodríguez}}
 \email{alexgu@fisica.uaz.edu.mx}
 \affiliation{Facultad de Física, Universidad Autónoma de Zacatecas, Apartado Postal C-580, 98060 Zacatecas, M\'exico}

\author{\firstname{T.}~\surname{Cisneros-Pérez} }
 \email{tzihue@gmail.com}
  \affiliation{Unidad Académica de Ciencias Químicas, Universidad Autónoma de  Zacatecas,Apartado Postal C-585, 98060 Zacatecas, M\'exico. }

 \author{\firstname{H.}~\surname{Garcia-Tecocoatzi}}
  \email{hugo.garcia.tecocoatzi@ge.infn.it}
 \affiliation{INFN, Sezione di Genova, Via Dodecaneso 33, 16146 Genova, Italy}

\author{\firstname{A.}~\surname{Dávila-Rivera}}
 \email{alejandra.davila@fisica.uaz.edu.mx}
\affiliation{Facultad de Física, Universidad Autónoma de Zacatecas, Apartado Postal C-580, 98060 Zacatecas, M\'exico}

\begin{abstract}
In this article, we explore machine learning techniques using support vector machines with two novel approaches: exotic and physics-informed support vector machines. Exotic support vector machines employ unconventional techniques such as genetic algorithms and boosting. Physics-informed support vector machines integrate the physics dynamics of a given high-energy physics process in a straightforward manner. The goal is to efficiently distinguish signal and background events in high-energy physics collision data. To test our algorithms, we perform computational experiments with simulated Drell-Yan events in proton-proton collisions. Our results highlight the superiority of the physics-informed support vector machines, emphasizing their potential in high-energy physics and promoting the inclusion of physics information in machine learning algorithms for future research.
\end{abstract}

\maketitle

\section{INTRODUCTION}

Machine learning techniques have proven to be extremely powerful when applied to high energy physics phenomena, both in theory and experimental studies~\cite{Whiteson2009, Komiske2019, Sharma2021}. Several algorithms have been applied to distinguish signals coming from high energy collider data~\cite{baldi2014, alves2017}. For instance, the discovery of the Higgs boson was aided with the help of the so-called boosted decision trees algorithm~\cite{biswas2023}. Some other popular machine learning algorithms which have been successful in high energy physics are: neural networks~\cite{Bhat2000, baldi2016, Auri2016, bishara}, linear regressions ~\cite{baldi2016, Murphy2018, Prosper} and deep learning~\cite{baldi2015, Chatham2020, barberio2017, Amacker2020}.

To continue exploiting the potential of machine learning techniques, the idea that physics insights can help design a better machine learning algorithm has recently been used across several fields, yielding excellent results. This field is known as physics-informed machine learning~\cite{karniadakis2021}. The majority of physics-informed machine learning studies are through the use of advanced neural network architectures. Moreover, support vector machines (SVM) which are based on kernel methods, have also benefited from these physics insights. The physics information in the SVMs is introduced via their kernels. The latter improves the SVMs performance~\cite{munduru2021}.

In the realm of high energy physics, physics-informed neural networks and deep learning techniques have been proposed to tackle the most challenging tasks in data analysis coming from high energy physics experiments ranging from searches of new physics phenomena to jet tagging~\cite{ngairangbam2024, congqiao2024, hao2023, Atkinson2022}. SVMs have been also helpful and interesting for the high energy physics community~\cite{sahin2016, atlasSVM2023, Vaiciulis2003, sforza2014, Whiteson2009, SauLanWu2021}. However, there is no reports of physics-informed support vector machines applied to high energy physics phenomena.  Hence, this invites the exploration of SVMs in the context of physics-informed machine learning.

This paper is focused on the application and interpretation of the SVMs in experimental high energy physics. The use of support vector machines is motivated by their relatively simple geometric interpretation, especially for binary discrimination of signal events against background events. First, we study what we call \textit{exotic} support vector machines. These SVMs are exotic in the sense that we utilize unconventional techniques to build them. That is, we use genetic and boosting algorithms to construct more efficient classifiers. Moreover, we use somewhat unconventional kernels. The construction of the exotic SVMs is guided by our previous studies~\cite{ramirez2022}. Second, we study physics-informed support vector machines. To include high energy physics information in our SVMs, we propose kernels that define the SVM and aim to capture the dynamical properties of the underlying theory that intends to describe the observed/expected data in high energy experiments. 

We perform a case of study: The \mbox{Drell-Yan} $Z$ boson production in proton-proton high energy collisions. In our studies, we simulate data for the process \mbox{$q\bar{q}\rightarrow Z \rightarrow l^+l^-$}, where $q$ and $\bar{q}$ are the quarks coming from the colliding protons and $l^+l^-$ are the final state oppositely charged leptons. Using the kinematic variables for these final state leptons we construct the kernels that define the SVM in every case. We then make formal statistical tests to compare the performances of each SVM. The latter will help us to conclude the usefulness of introducing the dynamics into a support vector machine algorithm.

In Sect.~\ref{method} we summarize the formalism of support vector machines, the basic kernel theory, and the definition of the considered kernels. Furthermore, we describe the genetic algorithms and boosting techniques used, and the approach of how to introduce the physics dynamics of a given process in high energy physics to a support vector machine algorithm. In Sec.~\ref{exp} we present the computational experiments to train and test our proposed support vector machines. In Sect.~\ref{results} we present our results and discussion. Finally, in Sect.~\ref{concl} we present our conclusions.

\section{Methodology}
\label{method}
We propose that if the theory underlying the dynamics of a physics process to be studied in high-energy experiments are considered or included during the construction of a kernel that defines the support vector machine, then the discrimination capabilities of the support vector machine binary classifier will be significantly enhanced. Then we compare the physics-informed SVMs with state-of-the-art SVMs. The following sections describe the ingredients of this proposal. 

\subsection{Support vector machines}
\label{svm}
In a binary SVM classifier, an optimal hyper-plane, separating two classes in the feature space, is found ~\cite{vapnik1995}. Binary classification is important in experimental high energy physics, as it helps discriminate between signals of interest against background. During optimization, the SVM model selects a subset of support vectors (SVs) from the training samples, $\mathbf{x}$, to establish the decision surface's location. To simplify the search for SVs, the training samples are mapped into a high-dimensional space using kernel functions, $\kappa(\mathbf{x}, \mathbf{z})$, which are expressed as inner products of the training samples or their mappings. In this feature space, a specific kernel produces a hyperplane that assigns a prediction $\mathbf{y}$ to each element of $\mathbf{x}$ based on which side of the hyperplane $\mathbf{x}$ lies. The kernel functions solve the optimization problem without explicitly using the actual mappings, a technique known as the kernel trick. Since data may not be perfectly separable and some points may lie within the margin or be misclassified, SVM implementations allow for a certain degree of misclassification by introducing an adjustable penalty cost $C$~\cite{vapnik1995, ramirez2022}. A SVM classifier is defined by its kernel and the
parameters that describe the kernel. Kernel theory in machine learning allows the construction of a broad diversity of kernels employing elemental kernel properties.
%\subsection{Kernel properties}
Let $\kappa_1$ and $\kappa_2$ be kernels over
$\mathbf{x} \otimes \mathbf{z},$ where $ \mathbf{x,z} \subseteq \mathbb{R}^n,$ 
$a \in \mathbb{R}^+$,
and $\kappa_3$ is a kernel over $\mathbb{R}^n \otimes \mathbb{R}^n$. 
Then the following functions are kernels as well~\cite{shawe2004}:
\begin{eqnarray}
\label{eq:kernProp}
\kappa(\mathbf{x}, \mathbf{z}) &=& \kappa_1(\mathbf{x},\mathbf{z}) + \kappa_2(\mathbf{x}, \mathbf{z})\\ \nonumber
\kappa(\mathbf{x}, \mathbf{z}) &=& a\kappa_1(\mathbf{x},\mathbf{z})\\\nonumber
\kappa(\mathbf{x}, \mathbf{z}) &=& \kappa_1(\mathbf{x}, \mathbf{z})\kappa_2(\mathbf{x}, \mathbf{z})\\ \nonumber
\kappa(\mathbf{x}, \mathbf{z}) &=& f(\mathbf{x})f(\mathbf{z})\\\nonumber
\kappa(\mathbf{x}, \mathbf{z}) &=& \kappa_3(\phi(\mathbf{x}),\phi(\mathbf{z}))\nonumber
\end{eqnarray}
where, $f,\phi: \mathbf{x} \rightarrow \mathbb{R}^n$.

\subsection{Basic kernels}
\label{kernel}

 In this context, a kernel is a Hermitian and positive semidefinite Gram matrix $G$ defined as $G = [ \langle v_j , v_i \rangle ]_{i,j=1}^n$, where the vectors $v_1, . . . , v_n$  live in a vector space that contains an inner product  $\langle \cdot ,\cdot \rangle$~\cite{roger2012}. To make the notation more compact, we write \mbox{$G = \kappa(\textbf{x}, \textbf{z})=\langle \textbf{x}, \textbf{z} \rangle$}, with $\mathbf{x,z} \subseteq \mathbb{R}^n$. This paper considers the kernels:
\begin{itemize}

    \item Linear kernel
    \begin{equation}
    \label{eq:lin}
    \kappa(\textbf{x}, \textbf{z})= \langle \textbf{x}, \textbf{z} \rangle,
    \end{equation}
     with no hyper-parameters. 
     
    \item  Radial Basis Function (RBF) kernel
    \begin{equation}
    \label{eq:lrbf}
    \kappa(\mathbf{x},\mathbf{z})= \exp (-\gamma\vert\vert \textbf{x} - \textbf{z}  \vert\vert^2 ),   
    \end{equation}
    with hyper-parameter $\gamma$.
    
    \item Sigmoid kernel
    \begin{equation}
    \label{eq:sig}
    \kappa(\textbf{x}, \textbf{z})=\tanh(\gamma\langle \textbf{x}, \textbf{z}\rangle+r),
    \end{equation}
    with hyper-parameters $\gamma$ and $r$.
    
    \item Polynomial kernel
    \begin{equation}
    \label{eq:pol}
    \kappa(\textbf{x}, \textbf{z})= (\gamma\langle \textbf{x}, \textbf{z}\rangle+r)^d,
    \end{equation}
    with hyper-parameters $\gamma$, $r$ and $d$.
     
\end{itemize}
%The hyper-parameters take values
For the sigmoid kernel $r=-1$. For the polynomial kernel $r=+1$ and $d=2$. Finally, we set a high $\gamma$ value, $\gamma = 100$, to provide a non-negligible impact of each training vector. The chosen values of the hyper-parameters $\gamma$, $r$, and $d$ enforce a good behavior when fitting a SVM~\cite{chang2010, lin2003}.

The kernels in Eqs.~(\ref{eq:lin})-(\ref{eq:pol}) in addition to the properties in Eq.~(\ref{eq:kernProp}) allow to define composed kernels with almost an arbitrary shape and hence help include the properties of a given physics process.

\subsection{Exotic support vector machines}
\label{exotic}
To construct exotic support vector machines we use and combine three elements:
\begin{itemize}
    \item Unconventional kernels. We use the kernels of Eqs.~(\ref{eq:lin})-(\ref{eq:pol}) arbitrarily joined according to Eq.~(\ref{eq:kernProp}). These kernels inherently do not carry any physical information beforehand.

    \item Ensembles of classifiers. An ensemble of classifiers is a collection of single weak classifiers that when combined together, provide a strong classifier~\cite{zhanga2012, sagi2018}. In this work, we use the AdaBoost algorithm~\cite{schapire1999} to construct ensembles. This adaptive method updates the vector\footnote{In this context, a vector is a point of the data sample.} weights based on the training error of a given binary classifier. These weights are used to train the next classifier to be added to the ensemble. Correctly classified vectors are assigned lower weights, whilst misclassified vectors are given higher weights. Thus, vectors that are harder to classify receive more focus from the algorithm. The AdaBoost algorithm is repeated $T$ times, $t=1,...,T$.  First, for the data true label $y_i$ and the base classifier prediction $h_t(\textbf{x}_i)$, the training error $\epsilon_t$ is calculated
    \begin{equation}
    \label{eq:err}
        \epsilon_{t}=\sum_{i=1}^{n}w_{i}^{t};\qquad y_{i}\neq h_{t}(\textbf{x}_{i}),
    \end{equation}
    where $w_i^t$ are the weights of each vector $\textbf{x}_i$ utilized to train the classifier. Then, the score $\alpha_{t}$ is defined as
    \begin{equation}
    \label{eq:alpha}
        \alpha_{t}=\frac{1}{2} \ln \frac{1-\epsilon_{t}}{\epsilon_{t}}\;.
    \end{equation}
    The weights are updated for the next iteration with
    \begin{equation}
        \label{eq:newW}
        w_{i}^{t+1}=w_{i}^{t}e^{ [  -\alpha_{t} y_{i} h_{t}(\textbf{x}_{i})]}\times A_{t},
    \end{equation}
    where $A_t$ is a normalization factor. The weights in Eq.~(\ref{eq:newW}) are applied to train and add a new classifier to the ensemble. When $T$ iterations are completed, the predicted label of the total ensemble is the weighted sum of the predictions of the individual classifiers within the ensemble
    \begin{eqnarray}
        \label{eq:final}
        H(\textbf{x}) = \sum_{t=1}^T \alpha_th_t(\textbf{x}).
    \end{eqnarray}

    \item Genetic algorithms. The genetic algorithms are optimization techniques inspired by the principles of biological evolution. Selections are performed using simple operators based on genetic recombinations and mutations. In this work, we use genetic algorithms to select a small subset of the training data, which will likely contain the support vectors needed to solve the binary classification problem~\cite{holland1992}. To determine if a subgroup of vectors is indeed likely to contain the support vectors, a fitness function is calculated to check if this subgroup is good at classifying data outside this subgroup. This is repeated for several subgroups of vectors and a selection of subgroups is performed using the high-low method~\cite{elamin2006}. The selected subgroups are recombined and the previous steps are repeated until a given stop criterion is satisfied. For more details, see Ref.~\cite{ramirez2022}.    
\end{itemize}

\subsection{The Drell-Yan process}
\label{dyproc}

Based on the parton model and the quark-antiquark annihilation mechanism, Sidney D. Drell and Tung-Mow Yan~\cite{drellYan1967} predicted the production of two oppositely charged leptons in hadron-hadron collisions. The neutral dilepton pair was predicted to appear with a large invariant mass. This production is the well-known neutral current Drell-Yan process. For proton-proton collisions, the partons participating in the Drell-Yan production are quark and antiquark that constitute the protons. The tree-level or leading-order partonic cross-section of the
\mbox{$q\bar{q}\rightarrow Z$} process is found to be~\cite{primer}
\begin{eqnarray}
    \hat{\sigma}^{q\bar{q}\rightarrow Z}&=& \frac{\pi }{3}
    \sqrt{2} G_F M_Z^2 (v_q^2+a_q^2) 
    \delta (\hat{s} -M_Z^2),
\label{eq:WZlo}
\end{eqnarray}
where $G_F$ is the Fermi weak coupling constant, $M_Z$ the invariant mass of the $Z$ boson, $v_q(a_q)$ is the
vector (axial vector) coupling of the $Z$ to the quarks, and $\hat{s}$ is the square of the center-of-mass energy of the quark-antiquark.

A quark with charge $Q_k$ inside a proton is described by a parton distribution function $q_k$. Considering all the proton parton distribution functions and with the aid of the QCD factorization theorem, it is found that the hadronic (proton-proton) cross-section for the Drell-Yan process is
\begin{eqnarray}
\label{eq:xSecZ}
 \frac{d \sigma^{pp \rightarrow Z}}{d M_Z^2} &=& \frac{\hat{\sigma}^{q\bar{q}\rightarrow Z}}{N_c} \int_0^1 {dx_1} {dx_2} 
 \delta(x_1 x_2 s -M_Z^2)\nonumber \\
& &\times \quad \Big[ \sum_k\;  Q_k^2\; \big( q_k(x_1,M_Z^2) \bar{q}_k(x_2,M_Z^2)\nonumber \\
& &  + \big[ 1 \leftrightarrow 2 \big]\big)  \Big],
\end{eqnarray}
where $1/N_c=1/3$ is the color factor. $x_{1,2}$ are defined in terms of the four-momentum of each parton
\begin{eqnarray}
    \label{eq:4mm}
    p_1^\mu &=& \dfrac{\sqrt{s}}{2}(x_1,0,0,x_1),\\
    \label{eq:5mm}
    p_2^\mu &=& \dfrac{\sqrt{s}}{2}(x_2,0,0,x_2).
\end{eqnarray}
From Eqs.~(\ref{eq:4mm})-(\ref{eq:5mm}) it is found that $\hat{s}=x_1x_2s$, where $s$ is the proton-proton center-of-mass energy. For the produced lepton pair, the rapidity is given by \mbox{$y= \textstyle{1/2 } \ln(x_1/x_2)$}, and hence  
\begin{equation}
\label{eq:eta}
    x_1 =\frac{M_Z}{\sqrt{s}}\;  \exp(y)\ ,\qquad x_2 =\frac{M_Z}{\sqrt{s}}\;  \exp(-y).
\end{equation}
\noindent
The cross-section of Eq.~(\ref{eq:xSecZ}) is multiplied by the branching ratio for any particular hadronic or leptonic final state of interest, which for this paper is the dielectron final state, namely, \mbox{$q\bar{q}\rightarrow Z\rightarrow e^+e^-$}.

The proton-proton cross section in Eq.~(\ref{eq:xSecZ}) is a function of the kinematics of the outgoing leptons. The kernel for our support vector machines is therefore constructed in accordance with Eqs.~(\ref{eq:WZlo})-(\ref{eq:eta}) in the following way: First, we identify the matrix of the proton-proton collision data as the kernel. Then, we perform operations on this kernel according to the relevant kinematic variables of the final state leptons in the cross-section. With this information, the kernel is expected to discriminate Drell-Yan events against backgrounds.
Taking into account the kernel properties in Eq.~(\ref{eq:kernProp}), we propose a physics-informed kernel
     \begin{equation}
     \label{eq:physker}
     \kappa(\mathbf{x}, \mathbf{z}) = \gamma(\langle \textbf{x}, \textbf{z} \rangle ^2 + \langle \textbf{x}, \textbf{z} \rangle + \langle \textbf{x}, \textbf{z} \rangle \cdot \exp (\langle \textbf{x}, \textbf{z} \rangle)).
    \end{equation}
    
The terms in Eq.~(\ref{eq:physker}) are intended to capture the physics in Eqs.~(\ref{eq:WZlo})-(\ref{eq:eta}) as
\begin{equation}
\label{eq:k1}
    \langle \textbf{x}, \textbf{z} \rangle ^2 \sim M_Z^2,
\end{equation}
\begin{equation}
\label{eq:k2}
    \langle \textbf{x}, \textbf{z} \rangle \sim M_Z,
\end{equation}
\begin{equation}
\label{eq:k3}
    \langle \textbf{x}, \textbf{z} \rangle \cdot \exp (\langle \textbf{x}, \textbf{z} \rangle) \sim \frac{M_Z}{\sqrt{s}}\cdot\exp(\pm y).
\end{equation}
In Eqs. (\ref{eq:k1})-(\ref{eq:k3}), when the $Z$ boson decays to an electron-positron pair, $M_Z$ is calculated from the kinematics of this electron-positron pair.

\section{Experiments}
\label{exp}

To test our proposed methodology, we perform computational experiments on a well-known Standard Model process. Namely, the production of a $Z$ boson decaying to an electron-positron pair (Drell-Yan production). Finally, we train, test, and compare several support vector machine binary classifiers to characterize their discrimination power between the Drell-Yan process against backgrounds.

\subsection{Data simulation}
\label{data}
\begin{figure}%[h!]
    \centering
    \includegraphics[scale=0.425]{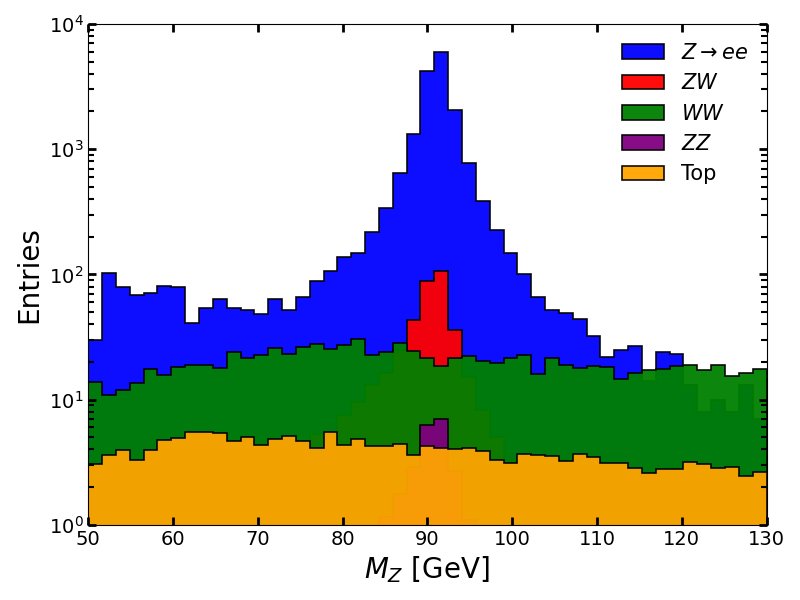}
    \caption{Invariant mass of the $Z$ boson calculated from the kinematics of the final state electron-positron pair coming from the simulated Drell-Yan events. The simulation was carried out with the \texttt{PYTHIA8.3} event generator~\cite{pythia8}.}
    \label{fig:Zboson}
\end{figure}

In this work, we consider Drell-Yan simulated signal and backgrounds. The simulated data is at the generator level, that is, no detector effects are taken into account. The simulation is carried out utilizing \texttt{PYTHIA8.3}~\cite{pythia8}. The event generation is performed utilizing the \texttt{PYTHIA} configuration for the production of weak single and double bosons for proton-proton collisions at center-of-mass energy $s=14$ TeV. For the signal events, we require that the event contains particles with the \texttt{PDGid}~\cite{Workman:2022ynf} corresponding to the $Z$ boson. Then, we require that this particle's invariant mass is within the $Z$ boson mass (91.1876 GeV) with a width of 40 GeV. Also, we require in the final state, two oppositely charged leptons whose mother particle is the selected $Z$. The kinematics of these charged leptons are the variables that are used to construct the kernels of the support vector machines. 
In this study, we consider the backgrounds which are most important for the Drell-Yan production $Z$ reported by the ATLAS and CMS experiments at the Large Hadron Collider~\cite{atlasZ2016, cmsZ2019}. The considered backgrounds are the diboson ($WW$, $ZW$, $ZZ$), $t\bar{t}$, and single top productions. These backgrounds are expected, as their final states may mimic the single $Z$ boson production final state charged leptons. The event selection for the backgrounds is similar to the single $Z$ boson. In this work, we do not consider backgrounds coming from multijet, as they are expected to be negligible ($<0.1\%$)~\cite{atlasZ2016, cmsZ2019}.
Since the events are simulated with no detector effects, the samples contain a high purity of events and there is no need to consider variables which are used to handle mismodelling, particle identification, lepton isolation, or acceptance effects. Figure~\ref{fig:Zboson} shows the invariant mass of the $Z$ boson calculated with the kinematics of the final state electron-positron pair.
Furthermore, we consider the electron-positron kinematic quantities: energy, momentum, transverse momentum, rapidity and azymuthal angle. These quantities are utilized to build the kernels for SVMs.
%The Drell-Yan $Z$ production along with the most important backgrounds are simulated with \texttt{PYTHIA8.3}.

\subsection{Data splitting}
\label{split}
In high energy physics, the challenge of class imbalance in the data sample usually appears. Hence, in this study, we consider different levels of imbalance among the signal and background events. Conventionally, in the binary classification task for high energy physics, a positive value is assigned to label a signal event, and a negative value is assigned to label a background event, being these values $\pm1$. We consider the cases when the data sample is fully balanced and the cases when there is an imbalance of the ratio signal: background as 1:3, 1:10, 3:1, and 10:1. This is summarized in Table~\ref{tab:samples}.

\begin{table}[htbp] 
%\centering
\caption{Drell-Yan data sets for the experiments.}
\label{tab:samples}
\begin{tabular}{c | c c c }
\toprule
Sample  & $+1$ Class & $-1$ Class &  Imbalance\\
\midrule

\texttt{half\_half}      & 5000      & 5000    & 1:1 \\
\texttt{1quart\_3quart}  & 2500      & 7500    & 1:3 \\
\texttt{3quart\_1quart}  & 7500      & 2500    & 3:1 \\
\texttt{1dec\_10dec}     & 1000      & 10000   & 1:10 \\
\texttt{10dec\_1dec}     & 10000     & 1000    & 10:1 \\
\bottomrule

\end{tabular}
\end{table}

\subsection{Support vector machine models}
The support vector machines we study in this paper are summarized in Table~\ref{tab:svms}. 
The models listed in this table are based on the definitions in Sections~\ref{svm}-\ref{dyproc}. The \texttt{phys-DY} model employs a kernel that incorporates the Drell-Yan dynamics, as detailed in Eqs.~(\ref{eq:WZlo})-(\ref{eq:eta}) and summarized in Eq.~(\ref{eq:physker}). Models with \texttt{lin}, \texttt{rbf}, \texttt{pol}, or \texttt{sig} in their names utilize the kernels specified in Eq.~(\ref{eq:lin}), Eq.~(\ref{eq:lrbf}), Eq.~(\ref{eq:pol}), and Eq.~(\ref{eq:sig}), respectively. Models featuring \texttt{adaboost} are ensembles constructed using the AdaBoost algorithm described in Sec.~\ref{exotic}, following Eqs.~(\ref{eq:err})-(\ref{eq:final}). Models marked with \texttt{gen} use genetic selection as discussed in Sec.~\ref{exotic}. Finally, \texttt{single} and \texttt{sum} indicate that the kernel consists of a single element or the sum of two kernels, respectively. In addition to the physics-informed support vector machine, the classifiers listed in this table are chosen for their outstanding performance in preliminary tests in agreement with our previous study in Ref.~\cite{ramirez2022}.

\begin{table}[htbp]
\caption{SVM models considered in this paper. The first column gives the name of the model, and the second provides a brief description of the elements considered to construct it.}
\label{tab:svms}
\centering
\resizebox{0.49 \textwidth}{!}{ 
\begin{tabular}{@{}ll@{}}
\toprule
Name & Description \\
\midrule
\texttt{phys-DY}            & Single with physics-informed kernel \\
\texttt{adaboost-gen-rbf}   & AdaBoost ensemble with genetic \\
                            & selection and RBF kernel \\
\texttt{adaboost-gen-pol}   & AdaBoost ensemble with genetic \\
                            & selection and polynomial kernel\\
\texttt{adaboost-gen-sig}   & AdaBoost ensemble with genetic \\
                            & selection and sigmoid kernel \\
\texttt{single-rbf}         & Single RBF kernel \\
\texttt{single-lin}         & Single linear kernel \\ 
\texttt{single-pol}         & Single polynomial kernel \\
\texttt{single-sig}         & Single sigmoid kernel \\
\texttt{single-sum-rbf-lin} & Sum of RBF and linear kernels \\
\texttt{single-sum-rbf-pol} & Sum of RBF and polynomial kernels \\
\texttt{adaboost-rbf}       & AdaBoost ensemble with RBF kernel\\
\texttt{adaboost-pol}       & AdaBoost ensemble with polynomial kernel\\
\texttt{adaboost-lin}       & AdaBoost ensemble with linear kernel \\
\texttt{adaboost-sig}       & AdaBoost ensemble with sigmoid kernel\\
\bottomrule
\end{tabular}}
\end{table}

%%%% RESULT PLOTS AND TABLES %%%%
% main plot
\begin{figure*}[!htbp]
     \centering
     \includegraphics[scale=.58]{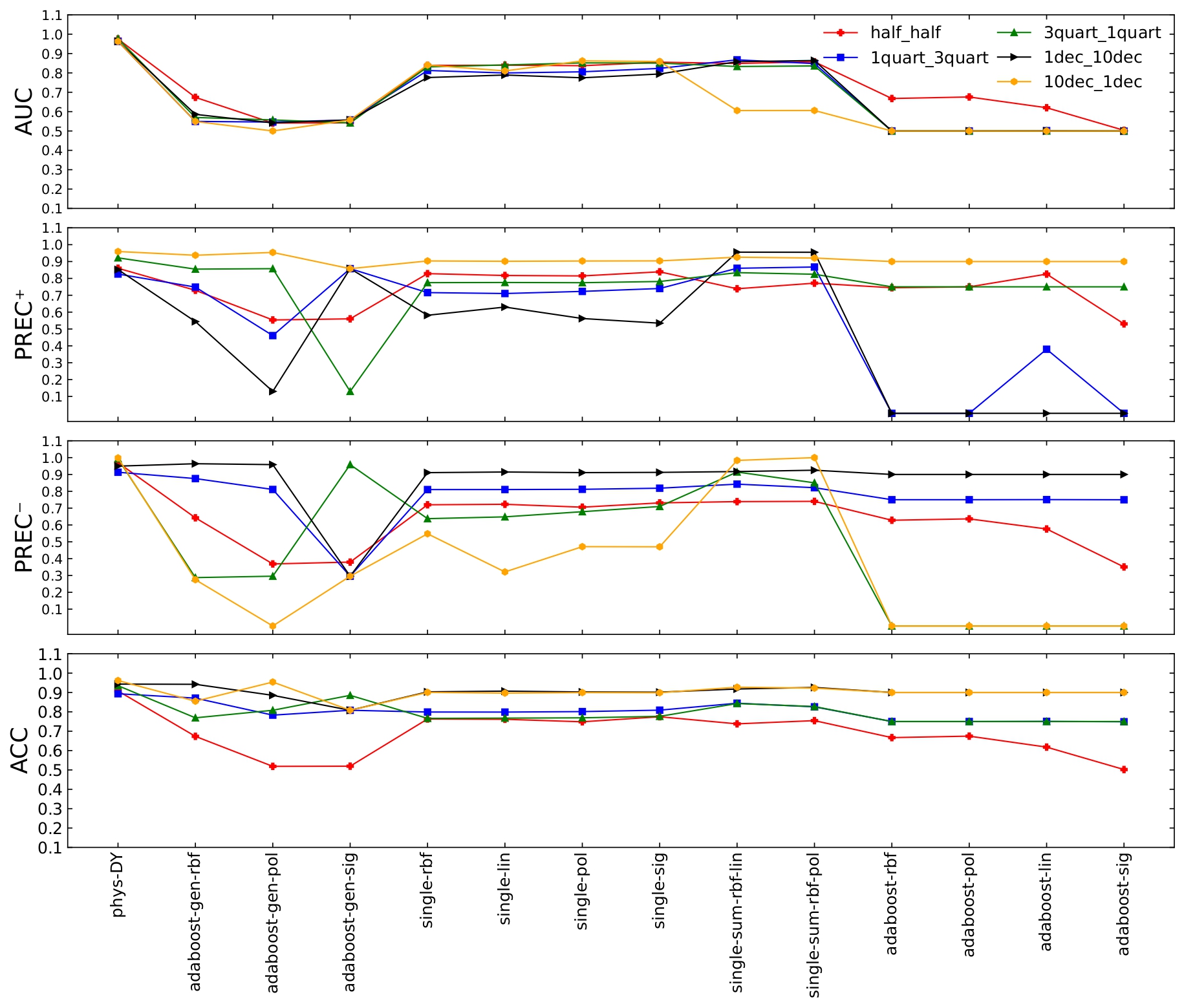}
     \caption{Performance metrics of the SVMs. The vertical axes correspond the average values of the ACC, PREC$^+$, PREC$^-$ and AUC as defined in Eqs.~(\ref{eq:acc})-(\ref{eq:prc-}). The error bars are not displayed. The horizontal axis indicates which SVM is being considered as listed in Table~\ref{tab:svms}. A solid line corresponds to a sample as listed in Table~\ref{tab:samples}.}
     \label{fig:results1}
\end{figure*}
% main table
\begin{table*}[htb!]
{\footnotesize
\begin{tabular}{c | c  c  c | c c c | c c c | c c c }\hline \hline
Model/Sample & $\mu_{AUC} (\sigma)$  & $p$-val.  &  R.$H_{0}$ & $\mu_{PREC^+}(\sigma)$  & $p$-val.  &  R.$H_{0}$ & $\mu_{PREC^-}(\sigma)$  & $p$-val.  &  R.$H_{0}$ & $\mu_{ACC}(\sigma)$  & $p$-val.  &  R.$H_{0}$ \\ 
\hline
\hline

\texttt{half\_half} &&&&&&&&&&\\
\texttt{phys-DY} & 0.98  (0.01) &&& 0.86 (0.02) &&& 0.97 (0.01) &&& 0.91 (0.01) \\
\hline
\texttt{adaboost-gen-rbf} &   0.67 (0.01)   & 0.0 & \checkmark\checkmark  &    0.73  (0.04)   & 0.0 & \checkmark\checkmark  &    0.64  (0.02)   & 0.0 & \checkmark\checkmark  &    0.67  (0.01)   & 0.0 & \checkmark\checkmark  \\
\texttt{single-rbf} &   0.84 (0.01)   & 0.0 & \checkmark\checkmark  &    0.83  (0.02)   & 0.0 & \checkmark\checkmark  &    0.72  (0.03)   & 0.0 & \checkmark\checkmark  &    0.76  (0.02)   & 0.0 & \checkmark\checkmark  \\
\texttt{single-sum-rbf-pol} &   0.86 (0.01)   & 0.0 & \checkmark\checkmark  &    0.77  (0.03)   & 0.0 & \checkmark\checkmark &    0.74  (0.03)   & 0.0 & \checkmark\checkmark  &    0.75  (0.02)   & 0.0 & \checkmark\checkmark  \\
\texttt{single-sum-rbf-lin} &   0.85 (0.01)   & 0.0 & \checkmark\checkmark  &    0.74  (0.03)   & 0.0 & \checkmark\checkmark  &    0.74  (0.03)   & 0.0 & \checkmark\checkmark  &    0.74  (0.02)   & 0.0 & \checkmark\checkmark  \\
\hline
\hline

\texttt{1quart\_3quart} &&&&&&&&&&\\
\texttt{phys-DY} & 0.96 (0.01) &&& 0.83 (0.04) &&& 0.91 (0.02) &&& 0.89 (0.01) \\
\hline

\texttt{adaboost-gen-rbf} &   0.55  (0.05)   & 0.0 & \checkmark\checkmark  &    0.75  (0.29)   & 0.36224 & \xmark  &    0.88  (0.02)   & 0.0 & \checkmark\checkmark  &    0.87  (0.01)   & 0.0 & \checkmark\checkmark  \\
\texttt{single-rbf} &   0.81  (0.02)   & 0.0 & \checkmark\checkmark  &    0.72  (0.06)   & 0.0 & \checkmark\checkmark  &    0.81  (0.02)   & 0.0 & \checkmark\checkmark  &    0.80  (0.02)   & 0.0 & \checkmark\checkmark  \\
\texttt{single-sum-rbf-pol} &   0.85  (0.02)   & 0.0 & \checkmark\checkmark  &    0.87  (0.05)   & 2e-05 & \checkmark  &    0.82  (0.02)   & 0.0 & \checkmark\checkmark  &    0.83  (0.01)   & 0.0 & \checkmark\checkmark  \\
\texttt{single-sum-rbf-lin} &   0.87  (0.02)   & 0.0 & \checkmark\checkmark  &    0.86  (0.05)   & 0.00025 & \checkmark  &    0.84  (0.02)   & 0.0 & \checkmark\checkmark  &    0.84  (0.01)   & 0.0 & \checkmark\checkmark  \\
\hline
\hline

\texttt{3quart\_1quart} &&&&&&&&&&\\
\texttt{phys-DY} & 0.98 (0.01) &&& 0.92 (0.01) &&& 0.99 (0.01) &&&  0.94 (0.01)\\
\hline

\texttt{adaboost-gen-rbf} &   0.57  (0.06)   & 0.0 & \checkmark\checkmark  &    0.85  (0.03)   & 0.0 & \checkmark\checkmark  &    0.29  (0.23)   & 0.0 & \checkmark\checkmark  &    0.77  (0.08)   & 0.0 & \checkmark\checkmark  \\
\texttt{single-rbf} &   0.83  (0.02)   & 0.0 & \checkmark\checkmark  &    0.77  (0.02)   & 0.0 & \checkmark\checkmark &    0.64  (0.08)   & 0.0 & \checkmark\checkmark  &    0.77  (0.02)   & 0.0 & \checkmark\checkmark  \\
\texttt{single-sum-rbf-pol} &   0.84  (0.02)   & 0.0 & \checkmark\checkmark  &    0.82  (0.02)   & 0.0 & \checkmark\checkmark  &    0.85  (0.04)   & 0.0 & \checkmark\checkmark  &    0.83  (0.02)   & 0.0 & \checkmark\checkmark  \\
\texttt{single-sum-rbf-lin} &   0.83  (0.02)   & 0.0 & \checkmark\checkmark  &    0.83  (0.02)   & 0.0 & \checkmark\checkmark  &    0.91  (0.04)   & 0.0 & \checkmark\checkmark  &    0.84  (0.01)   & 0.0 & \checkmark \checkmark \\

\hline
\hline

\texttt{1dec\_10dec}\\
\texttt{phys-DY} & 0.96 (0.01) &&& 0.85 (0.07) &&&  0.95 (0.01) &&& 0.94 (0.01)\\
\hline

\texttt{adaboost-gen-rbf} &   0.59  (0.06)   & 0.0 & \checkmark\checkmark  &    0.54  (0.33)   & 0.0 & \checkmark\checkmark  &    0.96  (0.01)   & 0.0 & \checkmark  &    0.94  (0.04)   & 0.59846 & \xmark  \\
\texttt{single-rbf} &   0.78  (0.03)   & 0.0 & \checkmark\checkmark  &    0.58  (0.15)   & 0.0 & \checkmark\checkmark  &    0.91  (0.01)   & 0.0 & \checkmark\checkmark  &    0.90  (0.01)   & 0.0 & \checkmark\checkmark  \\
\texttt{single-sum-rbf-pol} &   0.86  (0.02)   & 0.0 & \checkmark\checkmark  &    0.95  (0.06)   & 0.0 & \checkmark  &    0.93  (0.01)   & 0.0 & \checkmark\checkmark  &    0.93  (0.01)   & 0.0 & \checkmark\checkmark  \\
\texttt{single-sum-rbf-lin} &   0.86  (0.02)   & 0.0 & \checkmark\checkmark  &    0.96  (0.06)   & 0.0 & \checkmark  &    0.92  (0.01)   & 0.0 & \checkmark\checkmark  &    0.92  (0.01)   & 0.0 & \checkmark\checkmark  \\

\hline
\hline

\texttt{10dec\_1dec}\\
\texttt{phys-DY} & 0.96 (0.02) &&& 0.96 (0.01) &&&  0.99 (0.01) &&& 0.96 (0.01) \\
\hline

\texttt{adaboost-gen-rbf} &   0.55  (0.05)   & 0.0 & \checkmark\checkmark  &    0.94  (0.01)   & 0.0 & \checkmark\checkmark  &    0.27  (0.32)   & 0.0 & \checkmark\checkmark  &    0.86  (0.09)   & 0.0 & \checkmark\checkmark  \\
\texttt{single-rbf} &   0.84  (0.02)   & 0.0 & \checkmark\checkmark  &    0.90  (0.01)   & 0.0 & \checkmark\checkmark  &    0.55  (0.26)   & 0.0 & \checkmark\checkmark  &    0.90  (0.01)   & 0.0 & \checkmark\checkmark  \\
\texttt{single-sum-rbf-pol} &   0.61  (0.06)   & 0.0 & \checkmark\checkmark  &    0.92  (0.01)   & 0.0 & \checkmark\checkmark  &    1.00  (0.01)   & 0.17971 & \xmark  &   0.92  (0.01)   & 0.0 & \checkmark\checkmark  \\
\texttt{single-sum-rbf-lin} &   0.61  (0.06)   & 0.0 & \checkmark\checkmark  &    0.93  (0.01)   & 0.0 & \checkmark\checkmark &    0.98  (0.04)   & 0.01646 & \checkmark\checkmark  &    0.93  (0.01)   & 0.0 & \checkmark\checkmark \\

\hline \hline
\end{tabular}
\caption{The first column indicates the data sample and the model used to describe the data. The second column provides information on the AUC: first, the mean value $\mu_{AUC}$ is reported along with its uncertainty $\sigma$ in parentheses; then, the $p$-value from the Wilcoxon test is presented. This is followed by the result of rejecting the null hypothesis, $H_0$. A double \checkmark\checkmark\ indicates the rejection of $H_0$ and that \texttt{phys-DY} model performs better than the rest of the classifiers, while a single \checkmark\ indicates the rejection of $H_0$ and that \texttt{phys-DY} model performs worse than the rest of the classifiers. An \xmark\ indicates that $H_0$ cannot be rejected. The third column presents the same information as the second column but for PREC$^+$. The fourth column presents the same information as the second column but for PREC$^-$. The fifth column presents the same information as the second column but for ACC.
}
\label{tab:student}
}
\end{table*}
\subsection{Support vector machines training and testing}
\label{svm_train}
To evaluate the efficiency of the proposed support vector machines, we perform training and testing experiments utilizing the data described in Sec.~\ref{data}. In the training phase, a subset of the data is used to fit the model. During the testing phase, the fitted model obtains the predictions for the remaining data, where these predictions are the labels of whether a given data point is signal or background. To ensure reliable performance metrics for each support vector machine, we implement a repeated $k$-fold cross-validation. We divide the data into $k$ folds, where each fold is used once as the test set while the remaining $k-1$ folds serve as the training set. This process is repeated for each of the $k$ folds. That is, the entire $k$-fold cross-validation is repeated $N_{cv}$ times, with a different random split for each repetition. Overall, this results in $k\times N_{cv}$ training and testing cycles. The reported metrics are the average values of the obtained distributions, with one standard deviation as the associated errors~\cite{kuhn2013, ramirez2022}.

The classifier metrics are defined in terms of the error matrix elements: $TP$ is the number of true positive values, $TN$ is the number of true negative values, $FP$ is the number of false positive values, $FN$ is the number of false negative values~\cite{powers2008}. The metrics considered in this paper are the accuracy ACC,
\begin{equation}
\label{eq:acc}
    \text{ACC} = \frac{TP+TN}{TP+TN+FP+FN},
\end{equation}
the positive precision $\text{PRC}^+$,
\begin{equation}
\label{eq:prc+}
    \text{PRC}^+ = \frac{TP}{TP+FP},
\end{equation}
the negative precision $\text{PRC}^-$,
\begin{equation}
\label{eq:prc-}
    \text{PRC}^- = \frac{TN}{TN+FN},
\end{equation}
and the Area Under the Receiver Operating Characteristic Curve AUC. The AUC is the area under the plot of the $TP$ yields at different thresholds~\cite{bradley1997}. In SVMs, these thresholds are obtained by varying the offset of the hyperplane from the origin to produce different predictions. The values of these metrics are within the range \mbox{[0,1]} where 0 corresponds to the worst performance and 1 to the best performance.

\subsection{Computing implementation}
We work out our calculations on \texttt{Python}. In particular, we use the software \texttt{NumPy}~\cite{numpy2020} and the \texttt{libsvm}~\cite{libsvm2011} implementation of \texttt{scikit-learn}~\cite{scikit2011}. The computing framework for our experiments is publicly available on \texttt{GitHub}~\cite{github}.

\section{Results}
\label{results}

\subsection{Cross validation and statistical tests}
\label{stats}

In Fig.~\ref{fig:results1}, we display our ACC, PREC$^+$, PREC$^-$, and AUC defined in Eqs.~(\ref{eq:acc})-(\ref{eq:prc-}). The latter are calculated for the data samples listed in Table~\ref{tab:samples} and the support vector machines described in Table~\ref{tab:svms}. Each point in these plots represents the mean value of the metric calculated following the cross-validation procedure described in Sec.~\ref{svm_train}. Here we set $k=10$ and $N_{cv}=10$, that is, we compute 100 times the training and testing phases for each sample and support vector machine, and obtain a distribution for each metric. The displayed ACC, PREC$^+$, PREC$^-$, and AUC are calculated using the predicted classes from the test samples excluded during the training phase. In this plot, a line of a given color can show the behavior across all the proposed support vector machines for a specific data sample.

We carry out a comparison of the support vector machine containing the Drell-Yan kernel, against the support vector machines that showed the best four behaviors in Fig.~\ref{fig:results1} (we consider that the rest of the classifiers are evidently outperformed by our proposed physics-informed kernel). These are the \texttt{single-sum-rbf-pol}, \texttt{single-sum-rbf-lin}, \texttt{single-single-rbf},  and \texttt{adaboost-gen-rbf}
support vector machines whose kernels are described in Table~\ref{tab:svms}.
In this work, we use a paired ranked Wilcoxon test~\cite{wilcoxon1945}. This is equivalent to a Student's $t$-test for distributions with a non-Gaussian behavior. This test will determine if the difference between the metrics of the physics-informed support vector machine with respect to the others is statistically significant. Let $H_0$ be the null hypothesis that states that the metrics of the classifiers are equal. The purpose is to accept or reject $H_0$ in light of the distributions of the metrics obtained in the cross-validation procedure. We reject $H_0$ at a statistical significance level of $\alpha = 0.05$, meaning we conclude that the ACC, PREC$^+$, PREC$^-$, and AUC of two classifiers are indeed not equal if the $p$-value, coming from the Wilcoxon test, is below 0.05. Table~\ref{tab:student} summarizes these tests, for each metric we display the mean value of its distribution along with the associated error given by the standard deviation of this distribution. Moreover, in the column named R.$H_0$ we display the results of the Wilcoxon test: check marks, \checkmark or \checkmark\checkmark, indicate that the test rejects the null-hypothesis, and a cross mark, \xmark, indicates that $H_0$ is not rejected. Table~\ref{tab:student} contains the results for the samples described in Table~\ref{tab:samples}.

\subsection{Discussion}
The first feature to note from Fig.~\ref{fig:results1}, is that the values for ACC are stable across the different samples. The reason for this is that this metric takes an average of both the signal and background classification results. This metric is appropriate when describing a balanced data sample. A similar pattern is observed in the values found for the AUC. Conversely, large fluctuations arise when analyzing the signal precision, PRC$^+$, and the background precision PREC$^-$. From the plots in Fig.~\ref{fig:results1}, the most noticeable observation is when we look at the lines corresponding to the samples with imbalance 3:1 and 10:1. This poor behavior of most of the classifiers is expected since when there are not enough samples of one kind during the training phase, the support vector machine fails to describe both classes. 
Note that there could be a misleading assessment regarding a given classifier, as this classifier can achieve high AUC, ACC, and PREC$^+$, while the PREC$^-$ is near zero. Therefore, this suggests that the most important metrics are the positive and negative precisions. The latter implies that a good classifier is expected to be robust against imbalances in data samples, which is typically the case in high energy physics. 
Remarkably, our proposed physics-informed classifier \texttt{phys-DY} shows high values for all the metrics presented here. The reason for this could be that we have effectively captured the intrinsic properties of the data samples by incorporating physics information into the kernel of the support vector machine. Other classifiers also exhibit stable metrics across the samples, which can be explained by the fact that their kernels are similar to the one inspired by the Drell-Yan process.

From Table~\ref{tab:student}, we can quantitatively compare the physics-informed kernel against the best-performing kernels. The first notable feature is that, in most cases, we can reject $H_0$. This is evident as the \mbox{\checkmark or \checkmark\checkmark} appears in almost every case. Upon inspecting the metric values, when we reject $H_0$, there are two scenarios. First, the physics-informed kernel outperforms the exotic kernel, indicated by a double check mark (\checkmark\checkmark). Second, the exotic kernel outperforms the physics-informed kernel, indicated by a single check mark (\checkmark). In almost all of the metrics presented in this study, our proposed physics-informed kernel in Eq.~(\ref{eq:physker}) performs better than the other kernels. Specifically, when analyzing PRC$^+$, our physics-informed kernel performs excellently for the most imbalanced data samples, \texttt{1dec\_10dec}. PRC$^+$ is the metric that provides information about the performance of a classifier at finding signal events in the sample. Therefore, the PRC$^+$ attained by the physics-informed kernel demonstrates that this kernel is useful for high energy physics data. Moreover, the physics-informed kernel presents a stable PRC$^-$ when describing all the samples, demonstrating the robustness of this kernel against imbalance in data samples. There are two other kernels that show competitive metrics, namely, the \texttt{single-sum-rbf-pol} and \texttt{single-sum-rbf-lin} kernels. These kernels are the sums of the individual kernels defined in Eqs.~(\ref{eq:lrbf}) and (\ref{eq:pol}), and Eqs.~(\ref{eq:lrbf}) and (\ref{eq:lin}), respectively. When comparing them with the physics-informed kernel in Eq.~(\ref{eq:physker}), we conclude that these kernels can both capture the dynamical properties of the Drell-Yan cross-section.

\section{Conclusions}
\label{concl}
In this work, we analyze several types of kernels that define a support vector machine. A physics-informed kernel is proposed to describe simulated data of a simple and well-known Standard Model process. The physics of this process is introduced to the kernel in a simple and straightforward manner, by considering the functional form of the kinematic variables found in the cross-section and then transforming the matrix that represents the data according to these functional forms. To test the effectiveness of this method, we construct unconventional kernels that \textit{a priori} can overcome the typical challenges of high energy physics. We carry out statistical tests to determine if the physics-informed kernel is competitive compared to kernels constructed with sophisticated machine-learning algorithms. Remarkably, it is found that our proposed physics-informed kernel outperformed these algorithms. This finding motivates further investigation into the improvement of machine learning algorithms for more complex high energy physics data using the proposed approach. This simple method of introducing physics insights to kernel methods is proven to be effective, and since there is a connection between kernel methods and neural networks~\cite{wang2020}, the techniques we study in this paper can be extended to more modern machine learning algorithms based on kernel methods.\\

\begin{acknowledgments}
This work was funded by the CONAHCYT project I1200/311/2023.
T. C. P. thanks a CONAHCYT  postdoctoral fellowship.  A. G. R. thanks SNII (México).
\end{acknowledgments}

%\appendix
%\section{}
%\label{}

\end{document}